\documentclass[a4paper,fleqn,usenatbib]{mnras}

\usepackage{txfonts}

\usepackage[T1]{fontenc}
\usepackage{ae,aecompl}

\usepackage{graphicx}	
\usepackage{amssymb}	

\newcommand{\gta}{\gtrsim}

\title[NGC 253 disrupting dwarf]{Satellite accretion in action: a tidally disrupting dwarf spheroidal around the nearby spiral galaxy NGC 253}

\author[A. J. Romanowsky, D. Mart\'inez-Delgado et al.]
{Aaron J. Romanowsky,$^{1,2}$\thanks{E-mail: aaron.romanowsky@sjsu.edu (AJR)}
D. Mart\'inez-Delgado,$^{3}$
Nicolas F. Martin,$^{4,5}$
\newauthor 
Gustavo Morales,$^{3}$
Zachary G. Jennings,$^{6}$
R. Jay GaBany,$^{7}$
\newauthor Jean P. Brodie,$^{2,6}$
Eva K. Grebel,$^{3}$
Johannes Schedler,$^{8}$
and Michael Sidonio$^{9}$
\\
$^{1}$Department of Physics \& Astronomy, San Jos\'e State University, One Washington Square, San Jose, CA 95192, USA\\
$^{2}$University of California Observatories, 1156 High Street, Santa Cruz, CA 95064, USA\\
$^{3}$Astron.\ Rechen-Institut, Zentrum f\"ur Astronomie der Universit\"at Heidelberg, M\"onchhofstr. 12-14, 69120 Heidelberg, Germany\\
$^{4}$Observatoire astronomique de Strasbourg, Universit\'e de Strasbourg, CNRS, UMR 7550, 11 rue de l'Universit\'e, 67000 Strasbourg, France\\
$^{5}$Max-Planck-Institut f\"ur Astronomie, K\"onigstuhl 17, 69117  Heidelberg, Germany\\
$^{6}$Department of Astronomy and Astrophysics, University of California, 1156 High Street, Santa Cruz, CA 95064, USA\\
$^{7}$Black Bird Observatory II, Alder Springs, CA, USA\\
$^{8}$Chilean Advanced Robotic Telescope, Cerro Tololo Inter-American Observatory, Chile\\
$^{9}$Terroux Observatory, Canberra, Australia
}

\date{Accepted 2015-12-21. Received 2015-12-14; in original form 2015-11-06}

\pubyear{2016}

\begin{document}
\label{firstpage}
\pagerange{\pageref{firstpage}--\pageref{lastpage}}
\maketitle

\begin{abstract}
We report the discovery of NGC~253-dw2, a dwarf spheroidal (dSph) galaxy candidate
undergoing tidal disruption around a nearby spiral galaxy, NGC~253 in the Sculptor group:
the first such event identified beyond the Local Group.
The dwarf was found using small-aperture amateur telescopes, and followed up with 
Suprime-Cam on the 8\,m Subaru Telescope
in order to resolve its brightest stars.
Using $g$- and $R_{\rm c}$-band photometry, we
detect a red giant branch 
consistent with an old, metal-poor stellar population at a distance of $\sim$\,3.5\,Mpc.
From the distribution of likely member stars, we infer
a highly elongated shape with a semi-major axis half-light radius of $(2\pm0.4)$~kpc.
Star counts also yield a luminosity estimate of $\sim 2\times10^6 \, L_{\odot,V}$ ($M_V \sim -10.7$).
The morphological properties of NGC~253-dw2 mark it as distinct from
normal dSphs and imply ongoing disruption at a projected distance of $\sim$\,50\,kpc from the main galaxy.
Our observations support the hierarchical paradigm wherein 
massive galaxies continously accrete less massive ones,
and provide a new case study for dSph infall and dissolution dynamics.
We also note the continued efficacy of small telescopes for making big discoveries. 
\end{abstract}




\section{Introduction}

The modern paradigm of cold dark matter with a cosmological constant ($\Lambda$CDM)
predicts that galaxies form hierarchically -- growing through the gradual merging of
many smaller galaxies.
A giant spiral like our Milky Way is expected to undergo a succession of dwarf galaxy
accretion events which have different observational signatures,
depending on their occurrence in the past, present, or future.
Ancient accretion events can be detected through careful sifting of the chemo-dynamical
phase-space of halo stars.
Ongoing accretion is implied by the presence of satellite galaxies within the halo,
and imminent accretion is marked by the existence of field dwarfs near to their future hosts.

\begin{figure*}
		\includegraphics[width=8.35cm]{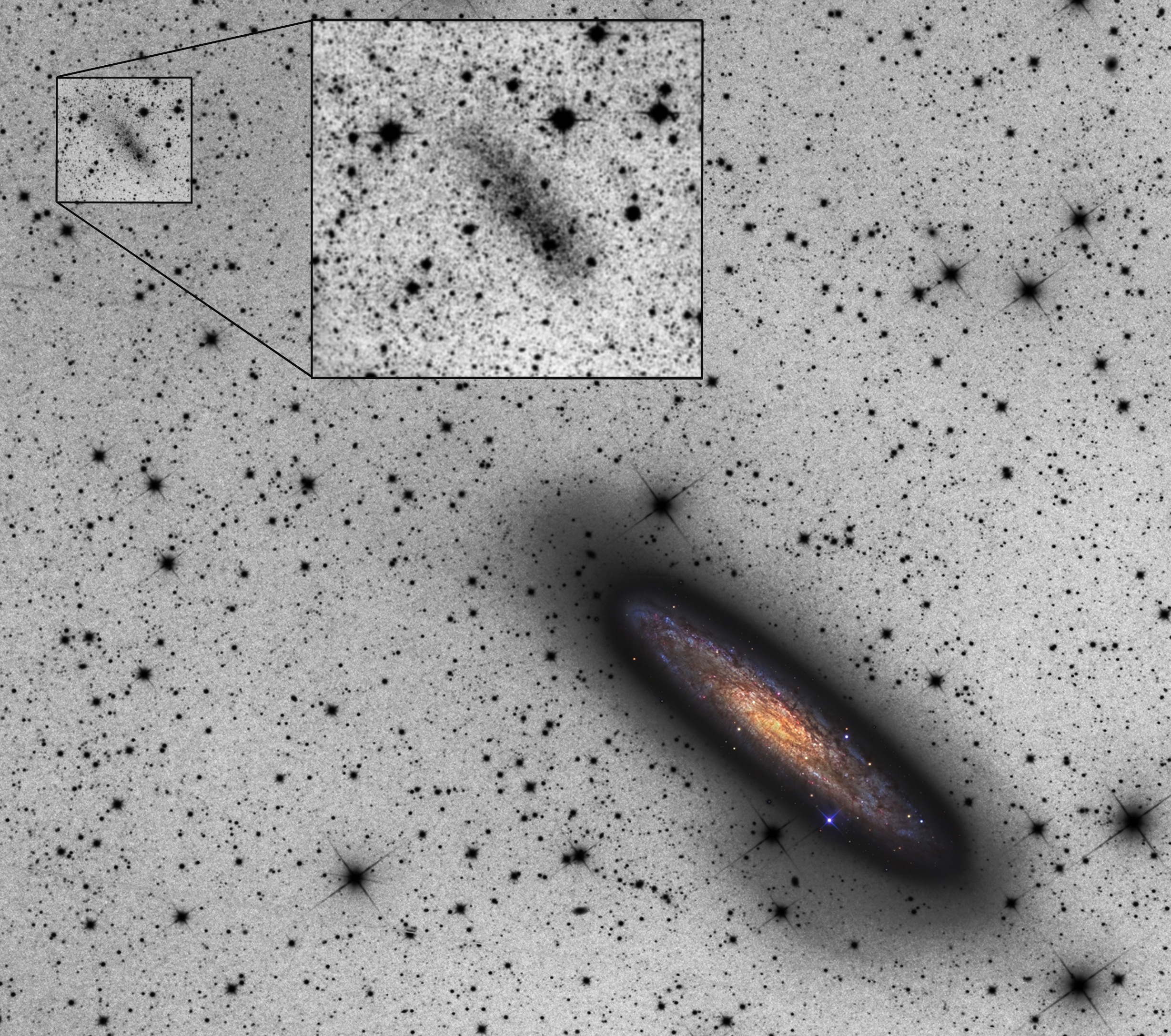} 
	\hskip 0.3cm
	\includegraphics[width=8.73cm]{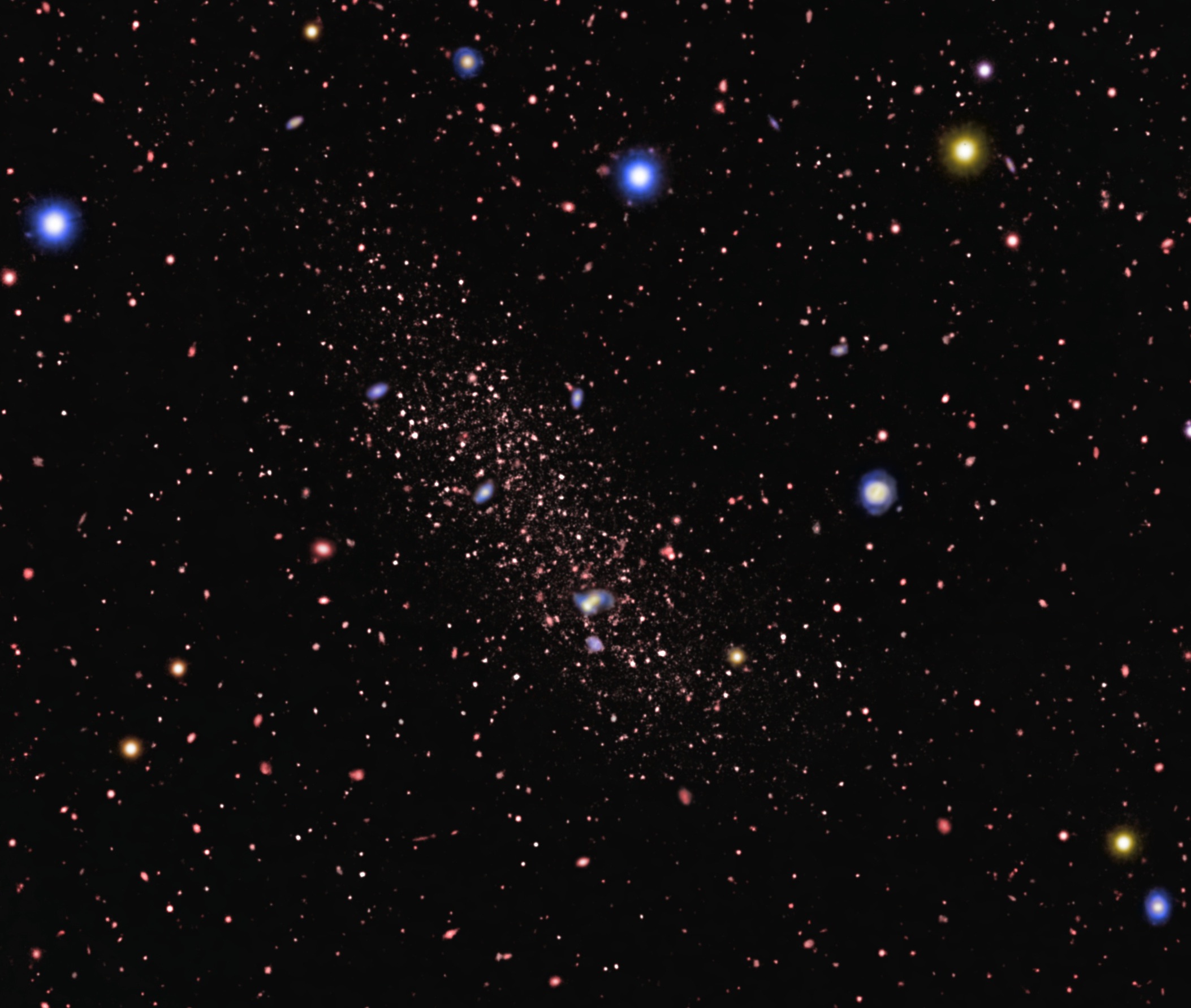} 
    \caption{{\it Left:} Amateur images of NGC~253 and its satellite NGC~253-dw2.
    The greyscale image is from 
    the 12-in telescope luminance filter, with colour image of NGC~253
    from the 0.5\,m BlackBird Remote Observatory,
    and field of view of 
    $\sim1\fdg1\times1\fdg0$
    ($67\times60$\,kpc).
    North is up and East is left.
    The zoom-in on NGC~253-dw2 is from CHART32 and covers
    $\sim8\times7$\,kpc.
    {\it Right:} Subaru/Suprime-Cam image, colourized using $g$ and $R_{\rm c}$ bands,
     and demonstrating that the satellite is
     composed of red, discrete stars.
    }
    \label{fig:bigimg}
\end{figure*}

All of these manifestations of accretion are  the focus of intense inventory and scrutiny,
in order to compare observations to theory.
In particular, 
there is a long-running concern about the relative scarcity of Milky Way satellites,
more recently expressed as a ``too big to fail'' problem (see summary in \citealt{Weinberg2015}).
There is also a recent controversy about the coherence of observed satellite systems compared to
expectations of more random infall (e.g. \citealt{Pawlowski2012,Ibata2013,Sawala2016}).
These tensions between observation and theory have led to doubts about the
standard model -- with solutions ranging from baryonic feedback to mild revisions of the dark matter theory
to radical dismissal of the entire cosmological framework
(e.g. \citealt{Kroupa2012,Brooks2014}).

In this context, intense observational efforts 
continue to focus 
on inventories and analyses of
satellites and streams around the Local Group (e.g. \citealt{Whiting2007,McConnachie2009,Bechtol2015}). 
More challenging but essential to a full picture is the extension
to studying
a broader sample of galaxies at larger distances.
Success has been obtained through 
low-surface brightness imaging
and resolved stellar maps
(e.g. \citealt{Merritt2014,Okamoto2015,Muller2015}), but the work is far from complete.

One valuable, alternative route to finding faint satellites and streams around nearby galaxies
is through using
small aperture (10--50 cm) telescopes in combination with the latest
generation of commercial CCD cameras
(e.g. \citealt{MD2012,MD2015,Karachentsev2015,Javanmardi2016}).
The short focal ratios of these telescopes,  along with the use of single, photographic-film size CCDs,
allow them
to probe large areas of galaxy haloes while
reaching surface brightness levels $\sim$2--3
magnitudes deeper ($\mu_{r} \sim$ 28 mag arcsec$^{-2}$) than the
classic photographic plate surveys (e.g. Palomar Observatory Sky Survey)
and the available large-scale digital surveys (e.g. SDSS).

Here we use this amateur telescope discovery approach to report
a faint, elongated galaxy projected onto the halo regions
of NGC 253 in the Sculptor Group.
With a distance of 3.5~Mpc, NGC~253  is one of the nearest large spirals, and has been well studied
(e.g. \citealt{Davidge2010,Bailin2011,Greggio2014,Monachesi2016}), 
but this object, which we call NGC~253-dw2, had been missed.\footnote{During the preparation of this paper, 
we learned of an independent discovery of the same object, using the 6.5\,m Magellan telescope \citep{Toloba2016}.}

\section{Observations}\label{sec:obs}

We first noticed a candidate satellite galaxy
in a visual inspection of several images of NGC~253 available on the internet that were
taken by amateur astronomers:
 Alessandro Maggi using a Takahashi Epsilon 180ED astrograph (18-cm 
diameter  at $f/2.8$) and Mike Sidonio using a 12-in $f/3.8$ Newtonian
(see Fig.~\ref{fig:bigimg}).
In both cases, the pixel scale was coarse and it was not clear if the elongated feature was real or
an artifact or reflection as often present in amateur images.  

We investigated the nature of NGC~253-dw2 by follow-up 
observations using amateur and professional facilities as follows.
First, deep imaging of the field centered on the dwarf candidate was collected remotely with the
Chilean Advanced Robotic Telescope (CHART32), an 80-cm $f/7$ corrected Cassegrain telescope
located at the Cerro Tololo Inter-American Observatory, Chile.
 An FLI PL-16803 CCD camera was used, with a pixel scale of 
 $0\farcs331$ 
 over a $22^\prime \times 22^\prime$ field of view. 
 Two overlapping sets of 20$\times$1200~s individual image frames were obtained through a Baader luminance filter 
 over several photometric nights between September and November 2013. 
 Each individual exposure was reduced following standard image processing procedures for dark subtraction, bias correction, and flat fielding 
\citep{MD2010}. 
The images were combined to create a final co-added image with a total exposure time of 48,000~s. 

Part of the resulting CHART32 image is shown as an inset in Fig.~\ref{fig:bigimg} (left panel).
The candidate object looks fairly regular and diffuse, like a nearby dwarf spheroidal (dSph) galaxy
rather than a background galaxy or an instrumental artifact.
It is also very elongated, and could in principle be a clump of foreground Galactic cirrus, 
although there are no such indications from the {\it Planck} 857~Ghz data.

\begin{figure*}
		\includegraphics[width=7.95cm,trim=2.7cm 0cm -0.6cm 0cm]{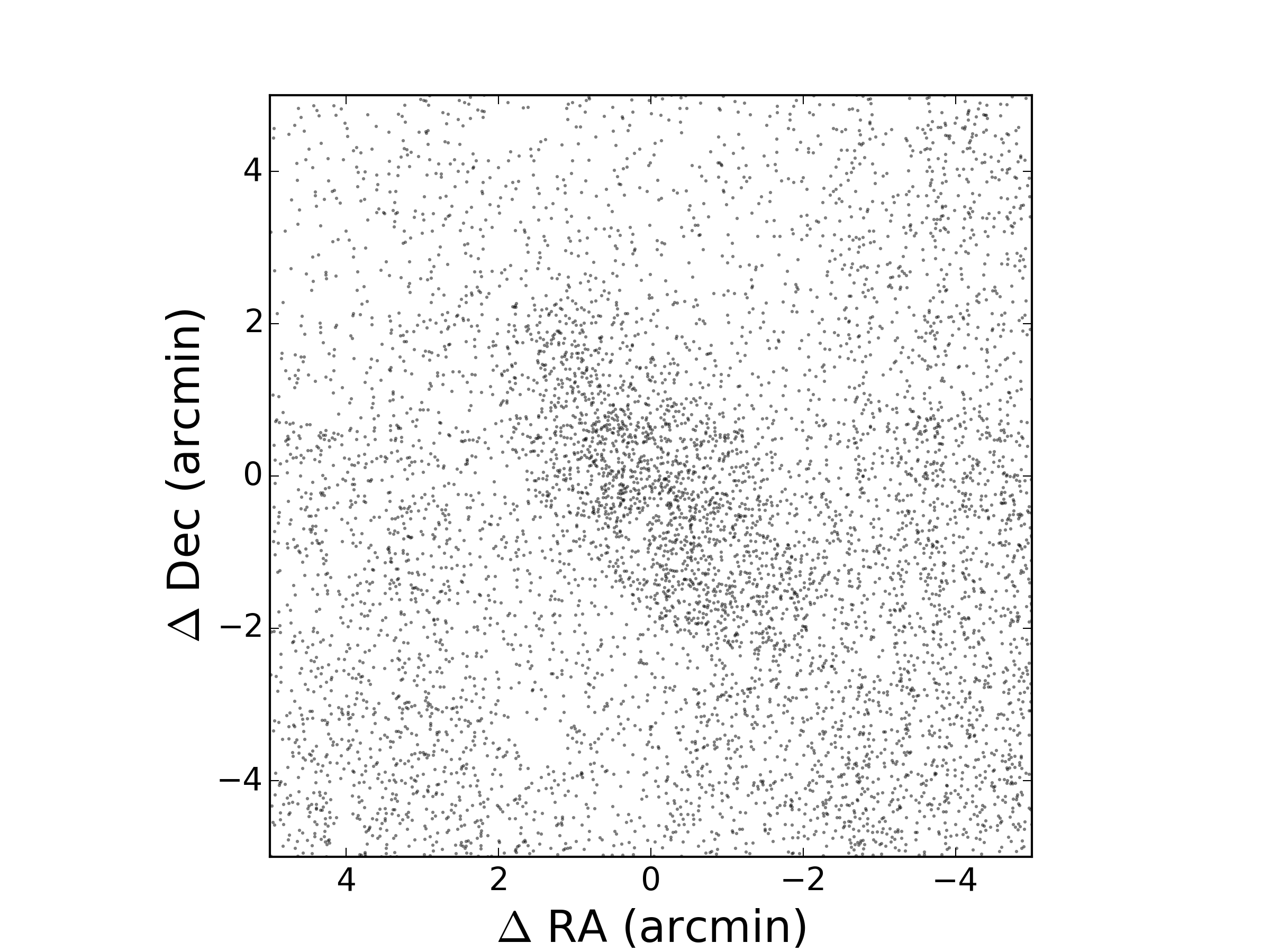}
		\includegraphics[width=9.66cm,trim=5.9cm 0.5cm 3.5cm 0cm]{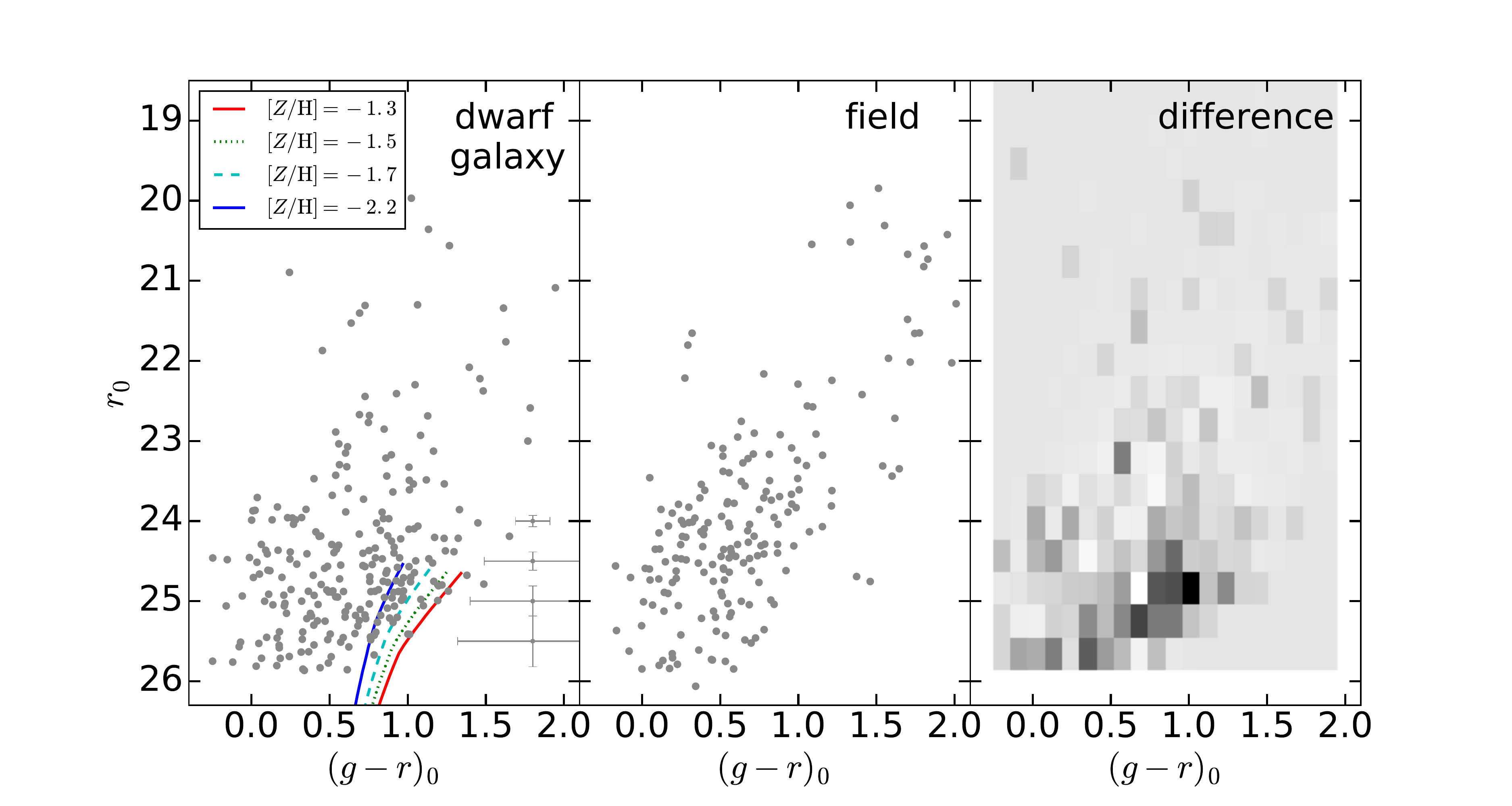}
    \caption{Resolved stars in NGC~253-dw2, from 
    Subaru/Suprime-Cam $R_{\rm c}$-band photometry.
    {\it Left:} star map in a $10^\prime$ (10\,kpc) square region.  
    The dwarf  galaxy emerges as an overdensity of stars against a uniform background of contaminants,
     and its elongated shape confirms the appearance in the amateur images (Fig.~\ref{fig:bigimg}, {\it left}).
    {\it Right:} colour--magnitude diagram for the dwarf ({\it left}), compared to an equal-area control field ({\it middle}).
    The binned difference between the dwarf field and a larger-area control field (for better statisics) is shown on the {\it right}.
   Curves show 13-Gyr isochrones from PARSEC \citep{Bressan2012}, with metallicities as in the legend.
    }
    \label{fig:resolved}
\end{figure*}

To investigate further, we used the Suprime-Cam imager on the 8\,m Subaru telescope
 \citep{Miyazaki2002},
with 5$\times$114\,s of $g$-band imaging 
on 2014 Dec 18, and 5$\times$120\,s of Johnson--Cousins
$R_{\rm c}$-band on 19 Dec 2014 (filters
dictated by observing schedule constraints).
The seeing was $\sim0\farcs8$ and $\sim0\farcs95$ in $g$ and $R_{\rm c}$, respectively.
The data were reduced using a modified version of the \texttt{SDFRED-2}
pipeline\footnote{\url{http://tinyurl.com/SDFRED2}}.

Preprocessing of the Suprime-Cam data was done by debiasing, trimming, flat fielding, and 
gain correcting 
each individual exposure chip-by-chip using median stacks of nightly sky flats.
Scattered light produced by bright stars both in and out of 
the field of view required
removing this smoothly varying component before performing photometry 
and solving for a World
Coordinate System solution. To remove scattered light, we fitted 
the smoothly varying component by creating a flat for every chip
 within each frame using
a running median with a 300 pixel box-size. This was
 then subtracted from the original,
unsmoothed frame to produce a final image for photometric processing

Fig.~\ref{fig:bigimg} (right panel) shows a portion of the Suprime-Cam image around NGC~253-dw2.
Again, the object appears elongated, but now is visibly resolved into stars -- with an appearance
similar to another Suprime-Cam image of a disrupted dSph at $\sim$\,4\,Mpc \citep{MD2012}.

The Suprime-Cam instrumental magnitudes were derived using the \textsc{daophot ii} and  \textsc{allstar} 
\citep{Stetson1987} packages. Sources
were detected in the $g$-band by requiring a 3-$\sigma$ excess above
the local background, with the list of the $g$-band detections  being 
subsequently used for the \textsc{allstar} run on the $R_{\rm c}$-band image. 
To obtain clean colour--magnitude diagrams and to
minimize contamination from faint background galaxies, 
 sources with the ratio estimator of the pixel-to-pixel scatter $\chi <$ 2.0 
and sharpness parameters
$|S| < 1.0$ \citep[see][]{Stetson1987} were kept for 
further analysis. 
To calibrate the photometry, we cross-identified sources from Subaru
with Pan-STARRS1 sources 
for which the photometric uncertainties are lower than 0.05
\citep{Tonry2012,Schlafly2012,Magnier2013}.
 There is no strong evidence of a significant colour term
between the Subaru and Pan-STARRS1 filters 
($g$ vs.\ $g$, $R_{\rm c}$ vs. $r$), so we used the median magnitude offset to
calibrate the Subaru data onto the Pan-STARRS1 photometric system.

\begin{figure*}
	\includegraphics[width=10.4cm,trim=1.9cm 9.0cm 7.4cm 0.0cm]{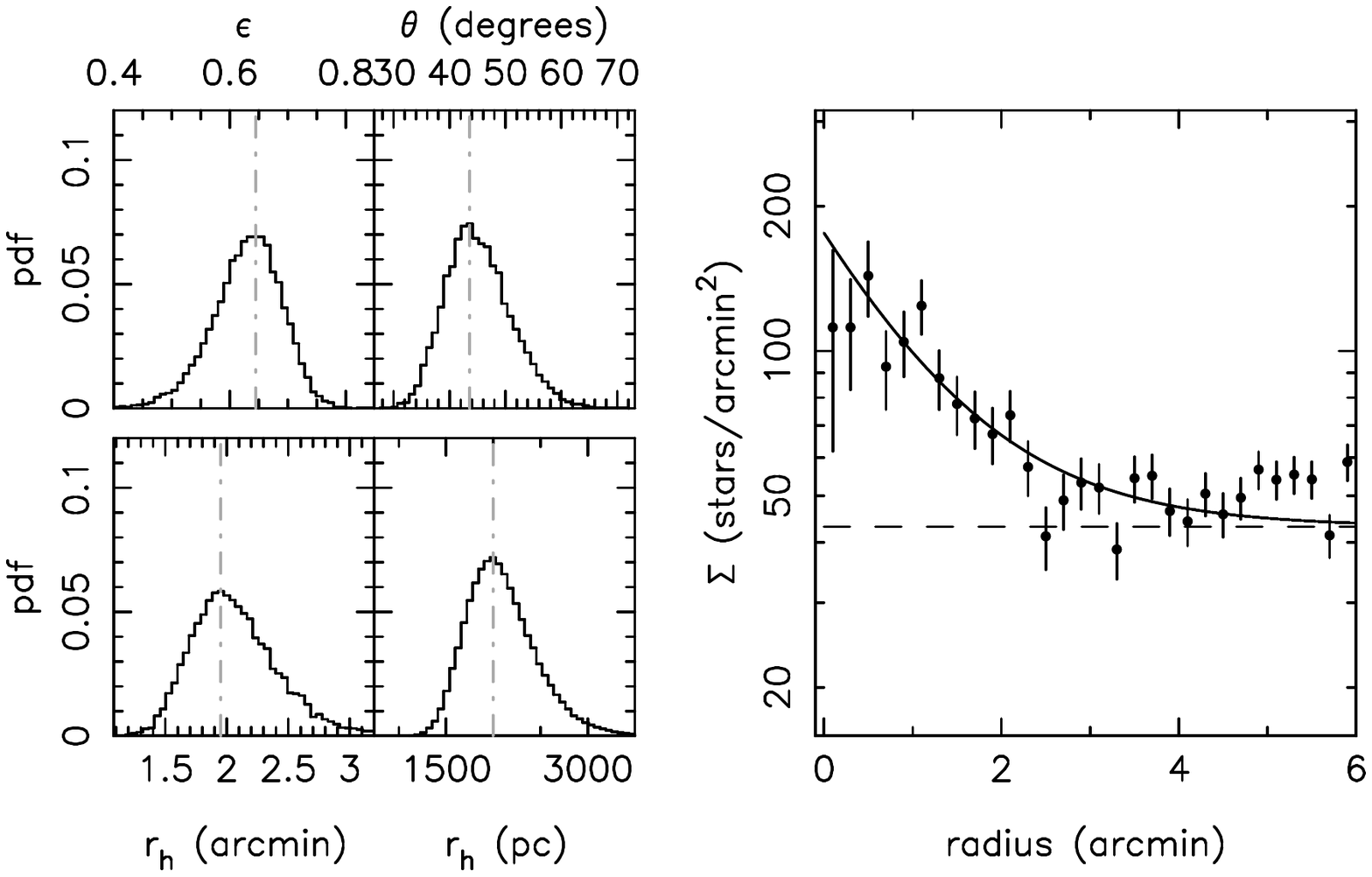}
		\includegraphics[width=7.2cm, trim=0.0cm 9cm 1.5cm 0.0cm]{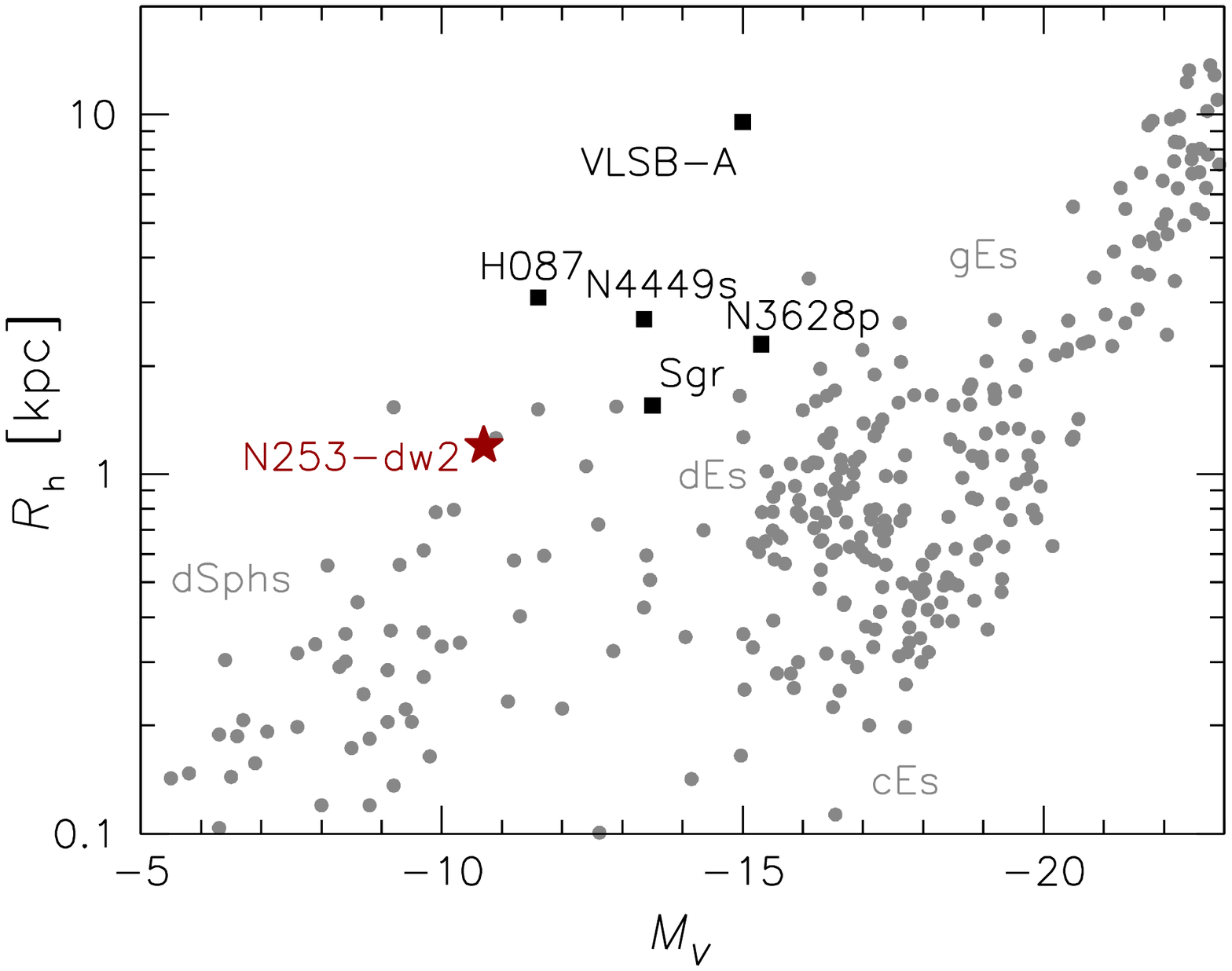}
    \caption{{\it Left:} 
    Structural parameters of NGC~253-dw2, from modelling star counts,
    and expressed as probability density functions.
    {\it Middle:} Model fit to
    star counts vs.\ semi-major axis radius;
    the dashed line marks the background level.
   {\it Right: } Circularized half-light radius vs.\ absolute magnitude for hot stellar systems.
NGC~253-dw2 ({\it red star}) has a large size for its luminosity.
Black squares mark other disrupting dwarfs
\citep{McConnachie2012,MD2012,Koch2012,Mihos2015,Jennings2015}.
    }
    \label{fig:PDFs}
\end{figure*}

\section{Colour--magnitude diagram and structural analysis}\label{sec:analysis}

We now consider some properties of NGC~253-dw2 using
 the Suprime-Cam resolved stellar photometry.
 Stars are selected from an elliptical region of $\sim 5^\prime\times1\farcm8$ around this galaxy
 (Fig.~\ref{fig:resolved}, {\it left panel})
and plotted on an extinction-corrected colour--magnitude diagram  ({\it right}).
For comparison, 
we also show a control field with an equal area but $\sim4^\prime$ away from the dwarf.
Relative to the control field, the dwarf field has an overdensity of stars
at roughly $r_0 \sim 25$, $(g-r)_0 \sim 1.0$.
This is just the signature expected for the upper regions of the red giant branch (RGB)
at a distance of $\sim$\,3.5\,Mpc -- based both on other observations of NGC~253 \citep{Sand2014}
and on theoretical isochrones (also shown in Fig.~\ref{fig:resolved}).

The presence of the RGB stars implies an age of more than 1~Gyr,
and the isochrone comparisons suggest a metallicity 
of [$Z$/H] $\sim -2.0 \pm 0.5$.
There are no obvious signs of blue main-sequence or of core-helium burning ``blue loop'' stars,
which would trace star formation over the past 600\,Myr.
The dwarf thus appears similar to the old, metal-poor halo population of NGC~253
(cf.\ \citealt{Radburn2011,Monachesi2016}),
although the 
shallowness of the imaging
makes it difficult to definitively rule out the presence of some young stars, or of
asymptotic giant branch stars from an intermediate-age population,
as found in another satellite of NGC~253, Scl-MM-Dw1 \citep{Sand2014}.
Therefore the possibility remains that it is a dSph--dIrr transition type.

The structural parameters of the dwarf galaxy are determined through modeling the spatial distribution of its stars,
assumed to follow an exponential density profile.
We use the algorithm developed in \citet{Martin2008} and updated
with a full Markov Chain Monte Carlo treatment. 
The posterior probability density function (PDF) is produced for six model parameters (plus a constant background level):
the  spatial centroid of the dwarf, its semi-major axis half-light radius $a_{\rm h}$, its ellipticity $\epsilon$, 
the major axis position angle
$\theta$, and the number of stars $N_\star$ in the galaxy
(down to a magnitude of $r_0=25.3$; the shallower $g$-band imaging is not used).

\begin{table}
	\centering
	\caption{Properties of NGC~253-dw2.}
	\label{tab:params}
	\begin{tabular}{cc} 
		\hline
		property & value \\
		\hline
		$\alpha$ (J2000) & 00:50:17.9 $^{+0.5}_{-0.6}$ \\ 
		$\delta$ (J2000) & $-$24:44:25.8 $^{+6.7}_{-8.4}$ \\ 
		ellipticity $\epsilon$ & $0.64^{+0.05}_{-0.07}$ \\
		position angle ($\deg$ E of N) & $44\pm6$ \\
		$M_V$ & $-10.7 \pm 0.4$ \\
		$a_{\rm h}$ (arcmin) & $2.0^{+0.4}_{-0.3}$ \\
		$a_{\rm h}$ (kpc) & $2.0\pm0.4$ \\
		$R_{\rm h}$ (kpc) & $1.1\pm0.2$ \\
		$\mu_{0,V}$ (mag arcsec$^{-2}$) & $26.2 \pm 0.8$ \\
		\hline
	\end{tabular}
\end{table}

The resulting marginalized PDFs  are in Fig.~\ref{fig:PDFs} ({\it left}) for the three main parameters of interest 
($\epsilon$, $\theta$ and $a_{\rm h}$) and summarised in Table~\ref{tab:params}. 
The new dwarf galaxy is very elongated: $\epsilon \sim 0.6$--0.7.
Its implied physical size is 
$a_{\rm h}= (2.0\pm0.4)$\,kpc
or $R_{\rm h} = (1.1\pm0.2)$\,kpc circularized
(geometric mean of major and minor axes;
including a small uncertainty on the distance modulus, $m-M = 27.7 \pm 0.1$, from an
adopted thickness of $\sim$\,150\,kpc for the NGC~253 satellite distribution).
Its total luminosity is estimated as follows. 
We begin populating an artificial stellar luminosity function, based on PARSEC isochrones,
until $N_\star$ stars fall above the observational threshhold of $r_0=25.3$.
We then sum up the flux of all stars, above and below the threshhold, and find a total 
$V$-band luminosity of $(1.6 \pm 0.3) \times 10^{6} L_\odot$,
which is equivalent to $M_V = -10.7 \pm 0.4$ and implies
a central surface brightness $\mu_{0,V}= 26.2 \pm 0.8$. 

\section{Discussion and conclusions} \label{sec:conc}

We now place NGC~253-dw2 in a plot of luminosity and size from a compilation of hot stellar systems
(\citealt{Brodie2011}, with updates\footnote{\url{http://sages.ucolick.org/spectral_database.html}}).
As Fig.~\ref{fig:PDFs} ({\it right}) shows, this object is about twice as large as
an average dSph of the same luminosity, although there are a few objects with comparable properties --
most notably And~XIX.
The elongated shape is also unusual for a dSph
\citep{McConnachie2012,Sand2012}, and the only known examples in the Milky Way with $\epsilon \geq 0.6$
are all thought to be tidally disrupting (Sgr, Her, UMa~I, UMa~II; e.g. \citealt{Munoz2010}).

Tides are predicted to cause elongation --
both from unbound stars soon after a pericentric passage
\citep{Penarrubia2009},
and during the final stage before complete disruption \citep{Munoz2008}.
We make a rough initial estimate of the tidal radius 
of NGC~253-dw2,  
assuming a $V$-band stellar mass-to-light ratio of $\sim$\,1.5, and the dark matter largely stripped.
At a galactocentric distance of 50\,kpc, we adopt a host galaxy 
circular velocity of 180\,km\,s$^{-1}$ (extrapolating from \citealt{Lucero2015}).
Following \citet{Munoz2008}, we predict 
a ``limiting'' radius of $\sim$\,1\,kpc,
which in combination with the observations
suggests the dwarf is filling or overfilling its Roche lobe.
Five examples of clearly disrupting dwarfs are also marked in  Fig.~\ref{fig:PDFs} ({\it right}), and
appear as similar outliers from the mean relation.
We conclude, based on the ellipticity and size of NGC~253-dw2,
that it is experiencing tidal disruption.
In this scenario, the ellipticity of the dwarf
may imply that $>90\%$ of its stellar mass has been lost \citep{Munoz2008},
and thus
that its pre-disruption luminosity was $\gta 2\times10^7 L_\odot$
(equivalent to the Fornax dSph).
If spread over a broad region, this large quantity of extra-tidal stars could well have been missed by our imaging.
Although such a luminous progenitor might be in tension with our low metallicity inference
(e.g. \citealt{Caldwell2006}),
we note that photometric errors make this inference very provisional.

The discovery of NGC~253-dw2, and the potential for follow-up observations, 
provides an opportunity to study the dynamics of a disrupting dSph, 
to test its dark matter distribution (e.g., \citealt{McGaugh2010,Errani2015})
and to compare its infall direction to the large-scale shear field \citep{Libeskind2015}.
It may also have implications for the effect of the satellite on the host.
Even though this dwarf has a relatively low stellar mass, 
$\Lambda$CDM models predict a pre-infall virial mass of
$\sim$(1--4)$\times10^{10} M_\odot$ \citep{Tollet2016}.
This is similar to the stellar mass of NGC~253 
($1.7\times10^{10} M_\odot$; \citealt{Lucero2015}),
and even if 95\% of the dark matter was lost before making a close passage to the host,
it could still have had a noticeable impact.
Intriguingly, NGC~253 has several peculiar features 
(central starburst, kinematic twisting, stellar halo ``shelf'' -- also visible in Fig.~\ref{fig:bigimg})
previously suggested as signatures of recent, unidentified interactions
(e.g. \citealt{Davidge2010}).
It is possible that the culprit has now been found, and that more cases of dwarf-galaxy ``stealth attacks''
could be identified through deep, wide-field studies of galaxy haloes.

\section*{Acknowledgements}

We thank the referee for helpful comments.
Based on data collected at Subaru Telescope (National Astronomical Observatory of Japan).
Funded by DFG Sonderforschungsbereich SFB 881 ``The Milky Way System" (subproject A2).
The Pan-STARRS1 Surveys have been made possible through contributions of the
Institute for Astronomy, the University of Hawaii, the Pan-STARRS Project
Office, the Max-Planck Society and its participating institutes, the Max
Planck Institute for Astronomy, Heidelberg and the Max Planck Institute for
Extraterrestrial Physics, Garching, the Johns Hopkins University, Durham
University, the University of Edinburgh, Queen's University Belfast, the
Harvard-Smithsonian Center for Astrophysics, the Las Cumbres Observatory
Global Telescope Network Incorporated, the National Central University of
Taiwan, the Space Telescope Science Institute, the National Aeronautics and
Space Administration under Grant No. NNX08AR22G issued through the Planetary
Science Division of the NASA Science Mission Directorate, the National
Science Foundation under Grant No. AST-1238877, the University of Maryland,
and Eotvos Lorand University (ELTE).








\bsp	
\label{lastpage}
\end{document}